\documentclass{elsart}

\usepackage{amssymb}

\begin{document}
\newtheorem{Corollary}{\normalsize C\scriptsize OROLLARY
\normalsize}

\begin{frontmatter}



\title{ Some sufficient conditions on Hamiltonian digraph}


\author{Guohun Zhu}
\address{Guilin University of Electronic Technology}
\address{No.1 Jinji Road, Guilin, China, 541004 }

\begin{abstract}
Z-mapping graph is a balanced bipartite graph $G$  of a digraph $D$ by split each vertex of $D$ into a pair of vertices of $G$. Based on the property of the $G$, it is proved that if $D$ is strong connected and $G$ is Hamiltonian, then $D$ is Hamiltonian. It is also proved if
$D$ is Hamiltonian, then $G$ contains at least a perfect
matching. Thus some existence
sufficient  conditions for Hamiltonian digraph and Hamiltonian graph are proved to be equivalent,  and two sufficient conditions of disjoint Hamiltonian digraph are given in this paper.
\end{abstract}

\begin{keyword}
Z-mapping \sep Hamiltonian cycle \sep balanced bipartite graph \sep
digraph

\end{keyword}

\end{frontmatter}

\section{Introduction}
It is well know that sufficient conditions for Hamiltonian cycle is
study in graph always precede the digraph \cite{Jorgen1997}. For
instance the famous Dirac condition is presented in 1952.
\begin{thm}
\cite{Dirac1952} \label{dirac1952}
 Let $G$ be a graph with $n > 2$ vertices. If $d(u) \geq n/2$
for every vertices, then $G$ is Hamiltonian.
\end{thm}

The similar sufficient condition for digraph is presented in 1960.
\begin{thm}
\cite{Ghouila1960}
\label{ghouila1960}
 Let $D$ be a strong connected digraph with $n > 2$ vertices. If $d(u) \geq n$ for every vertices, then $D$ is Hamiltonian.
\end{thm}

It is not difficult to find that the degree in digraph is large
twice than in undirected graph. Thus, we can think that the
sufficient conditions of Hamiltonian digraph and the Hamiltonian
graph have some relationship.

In this paper, A relationship  is  discover by a mapping between
digraph and graph, which is named as Z-mapping graph \cite{zhu2008} . The first this term
is called {\it "projector graph" } when restrict the digraph with
degree bound two \cite{zhu2007}. Based on this bijection, it is easy that some pervious sufficient conditions for Hamiltonian digraph and Hamiltonian graph are equivalent in the degree. And two disjoint Hamiltonian digraph is easy determined from the disjoint perfect matching bipartite graph.

Throughout the paper, we stick to the graph  terminology and
notation as given in \cite{Harary1969} and \cite{zhu2008} and focus on only finite simple (un)directed graph, i.e. the graph has no
multi-arcs and no self loops. However,
it need to recapitulate the cycle definition in graph and digraph
with matrix forms respectively.

\begin{defn}
A {\it simple cycle } $L$ in digraph is a set of arcs
$(a_{1},a_{2},\ldots,a_{l})$, the incidence matrix $C$ could be
permutated by row or column exchange to following forms

  \begin{equation}
  \label{dcycle}
    C = \left(
        {\begin{array}{llllll}
            1&0&0&\ldots&0&-1\\
            -1&1&0&\ldots&0&0\\
            0&-1&1&\ldots&0&0\\
            0&0&-1&\ldots&0&0\\
            0&0&0&\ldots&1&0\\
            0&0&0&\ldots&-1&1
        \end{array} } \right ).
\end{equation}

\end{defn}

\begin{defn}
A {\it simple cycle } $L$ in graph is a set of edges
$(e_{1},e_{2},\ldots,e_{l})$, the incidence matrix $C$ could be
permutate by row col exchange to following forms
  \begin{equation}
          C = \left(
        {\begin{array}{llllll}
            1&0&0&\ldots&0&1\\
            1&1&0&\ldots&0&0\\
            0&1&1&\ldots&0&0\\
            0&0&1&\ldots&0&0\\
            0&0&0&\ldots&1&0\\
            0&0&0&\ldots&1&1
        \end{array} } \right ).
\end{equation}

\end{defn}

The cycle definition responses to an  edge set of $G$, which is easy
to build a mapping from matching which is another definition on an
edge set of  graph $G$.

\section{Z-mapping}
Firstly, let us recapitulate the incidence matrix divided method in
\cite{zhu2008}, which divided the incidence matrix $C$ of a digraph into two groups.
\begin{equation}
\label{C_Plusedef}
      C^+=\left\{c_{ij} | c_{ij} \geq 0 \mbox{ otherwise  $0$ }\right \}
\end{equation}
\begin{equation}
\label{C_Minusdef}
      C^-=\left\{c_{ij} | c_{ij} \leq 0  \mbox{ otherwise  $0$ } \right \}
\end{equation}

Secondly, let us combine these two matrices into a matrix.
\begin{defn}
\label{incidencematrixdef} Let $C$ be a incidence matrix of digraph
$D$, the Z-mapping graph of $D$, denoted as $Z(D)$, is defined as a
balanced bipartite graph $G(X,Y:E)$ with a incidence matrix $F=\left
(
    {\begin{array}{c c}
    C^+  \\
    -C^-
    \end{array}}\right)$,
\end{defn}

In fact, it is a isomorphism build from $D(V,A)$ to $G(X,Y;E)$. Let the $V^+=\{v_{i}^{+}\}$ represents the vertices in $C^+$ and the $V^-=\{v_{i}^{-}\}$ represents the vertices in $C^-$. More precisely, the mapping  $D(V,A) \rightarrow G(V^+,V^-;E)$ defines a map from arcs of $D$ to edges of $G$:
\begin{equation}
\label{arc2edge}
E=\{ (v_{i}^{+},v_{j}^{-}) | <v_{i},v_{j}> \in A \}.
\end{equation}

Based on the same principle, we also define a reverse mapping from a simple balance bipartite graph $G$ to a simple digraph $D$.
\begin{equation}
f^{-1}: G(V^+,V^-;E) \rightarrow D(V;A)
\end{equation}
this function is a reverse function of the equation~\ref{arc2edge}:
\begin{equation}
\label{edge2arc}
A=\{ <x,y> | (x,y) \in E \wedge x \in V^+ \wedge y \in V^- \}.
\end{equation}

Thus first main result is follows:
\begin{thm}
\label{digraph2graph}
 Given a Z-mapping graph $G$ of a strong connected digraph
$D$, if $G$ is Hamiltonian, then $D$ is Hamiltonian.
\end{thm}

\begin{pf}
Suppose the $G$ has a Hamiltonian cycle $L$ and the response
incidence matrix is a $2n \times 2n$ matrix $C$.
  \begin{equation}
          C = \left(
        {\begin{array}{llllll}
            1&0&0&\ldots&0&1\\
            1&1&0&\ldots&0&0\\
            0&1&1&\ldots&0&0\\
            0&0&1&\ldots&0&0\\
            0&0&0&\ldots&1&0\\
            0&0&0&\ldots&1&1
        \end{array} } \right ).
\end{equation}
And divide $C$ to  two groups: $C_1$ and $C_2$, where
$$
C_1=\left\{c_{i} | \mbox {i is odd}  \right \}
$$
$$
C_2=\left\{c_{i} | \mbox {i is even} \right \}
$$

Let $C^\prime=C_1 - C_2$. Then obtain a matrix $n \times 2n$.
  \begin{equation}
          C^\prime = \left(
        {\begin{array}{llllll}
            0&-1&0&\ldots&0&1\\
            0&1&0&\ldots&0&0\\
            0&0&0&\ldots&0&-1
        \end{array} } \right ).
\end{equation}

After the column from $C^\prime$ with all zero is removed. Then a
new matrix $n \times n$ is obtained. This matrix satisfies the cycle
equation~\ref{dcycle}.
\end{pf}

The second main result is follows:
\begin{thm}
\label{graph2digraph} Given a Z-mapping graph $G$ of a  digraph $D$,
if $D$ contains a  Hamiltonian cycle $L$, then $G$ contains a
perfect matching $M$, where $M=F(L)$.
\end{thm}
\begin{pf}
Suppose that $D$ has a Hamiltonian cycle $L$, then $D$ has a sub
incidence matrix as equation~\ref{dcycle}. According to the divided
matrices approach, the response Z-mapping graph $G(X,Y;E)$ is a
1-regular bipartite graph, thus $\forall x \in X \; |N(x)|=|X|$,
thus according to the Hall \cite{Hall1935}rules, the graph z-mapping
graph $G$ is a perfect matching.

\end{pf}

\section{Sufficient conditions from Hamiltonian  Graphs to Digraphs}
According to the theorem~\ref{digraph2graph}, the
theorem~\ref{ghouila1960} can be directly generated from
theorem~\ref{dirac1952}. As the Dirac theorem has extended to the $d(u)<n/2$ case by the Faudree etc. as follows.
\begin{thm}
\cite{Faudree2006}
\label{dirac2006}
 Let $G(V,E)$ be a graph with $n > 2$ vertices and a subset $S=\{v : d(v) <\frac{n}{2}\}$ of $V$, if $k$ is the minimum degree of $G$, and $|S| \leq k-1$  then $G$ is Hamiltonian.
\end{thm}

Thus as the theorem~\ref{digraph2graph}, it can be derived to a new
theorem.
\begin{thm}
\label{mytheorem3}
 Let $D$ be a strong connected digraph with $n > 2$ vertices and $S=\{v : d(v) <n \}$ . if $k$ is the minimum degree of $D$, and $|S| \leq k-1$, then
 $D$ is Hamiltonian.
\end{thm}

In fact, Moon and Moser  \cite{Moon1963} have presents a theorem for similar sufficient condition on bipartite Hamiltonian graph.
\begin{thm}
\label{Moon_less_n}
\cite{Moon1963}
Let  $B$ be $2n$ balance bipartite graph, where 1 <k<n, if the number of vertices $S=\{v: d(v)<k \}$  $k<n$ is less than $n$, then $B$ is Hamiltonian.
\end{thm}

In \cite{Moon1963}, Moon and Moser also presented a theorem for
balanced bipartite graph as follows.
\begin{thm}
 \cite{Moon1963}
 \label{no_use1}Given a balanced bipartite graph
with $2n>2$ vertices, for all vertices if $d(u)>\frac{n}{2}$, then $G$ is Hamiltonian.
\end{thm}

It is easy to find that the theorem~\ref{no_use1} could derive a sufficient condition  $d^+(u)+d^-(u)=d(u)>n$ for Hamiltonian
digraph, but this is not precede the theorem~\ref{ghouila1960}.

Since a Hamiltonian cycle consists of two independent perfect
matching in balanced bipartite graph. Thus a sufficient condition
for Hamiltonian digraph as follows.
\begin{Corollary}
\label{no_use} Given a strong connected digraph with $n > 2$
vertices. for all vertices if $d^+(u)>\frac{n}{2}$ and
$d^-(u)>\frac{n}{2}$ , then $D$ contains two disjoin Hamiltonian
cycle.
\end{Corollary}


\section{Sufficient conditions from Hamiltonian digraph  to perfect bipartite graph}
The relation perfect matching and cycle .
For instance, in 1972, Las Verganas obtained the following results:
\begin{thm}
\cite{Verganas1972}  Let $G(X,Y;E)$ be a balanced bipartite graph
with $2n > 2$ vertices. For  all vertices $u \in X$, $v \in Y$ ,
$(u,v) \not \in E$, If $d(u) +d(v)  \geq  n+2$, then every perfect
matching in $G$ is contained in a Hamiltonian cycle.
\end{thm}

According to theorem~\ref{digraph2graph}, it could derived a
sufficient condition for Hamiltonian digraph from Ore conditions
\cite{Ore1960} such as  $d(u)+d(v) \geq n$, but Woodall had obtained a
more strict sufficient condition.
\begin{thm}
\cite{woodall1960} Let $D$ be a strong connected digraph with $n >
2$ vertices. For all vertices $u, v , <u,v> \not \in A$,  If
$d^+(u)+d^-(v) \geq n$, then $D$ is Hamiltonian.
\end{thm}

Combination these two theorems, we could get two interesting
results.
\begin{Corollary}
Let $D(V,A)$ be a digraph with $n > 2$ vertices. For  all vertices
$u,v$ , $<u,v> \not \in A$, If $d^+(u) +d^-(v)\geq n+2$, then $D$
contains at least two disjoin Hamiltonian cycles.
\end{Corollary}

\begin{Corollary}
Let $G(X,Y;E)$ be a balanced bipartite graph with $2n > 2$ vertices.
For  all vertices $u \in X$, $v \in Y$ , $(u,v) \not \in E$, If
$d(u)+d(v) \geq n$, then $G$ contains at least one perfect matching, if
If $d(u) +d(v) \geq n+2$, then $G$ contains at least two disjoin
perfect matchings.
\end{Corollary}




\end{document}